\documentclass[pdftex,epjc3,twocolumn]{svjour3}          

\RequirePackage[T1]{fontenc}
\usepackage{amsmath}
\usepackage{amssymb}

\smartqed  
\usepackage{bm}
\usepackage{subfig}
\usepackage{multirow}
\RequirePackage{graphicx}
\RequirePackage{flushend}
\RequirePackage[numbers,sort&compress]{natbib}
\RequirePackage[colorlinks,citecolor=blue,urlcolor=blue,linkcolor=blue]{hyperref}
 
\def\be{\begin{equation}}
\def\ee#1{\label{#1}\end{equation}}
\newcommand{\ben}{\begin{eqnarray}}
\newcommand{\een}{\end{eqnarray}}

\begin{document}

\title{ECO signatures from the Weyl-induced gravitational waveform injection} 
\author{Abra\~ao J. S. Capistrano\thanksref{e1,addr1,addr2}
\and
        Newton M. S. Ch\'avez\thanksref{e2,addr1} 
\and
V\'ictor A. M. Le\'on\thanksref{e3,addr1}
}

\thankstext{e1}{e-mail:capistrano@ufpr.br}
\thankstext{e2}{e-mail:newton.chavez@unila.edu.br}
\thankstext{e3}{e-mail:victor.leon@unila.edu.br}

\institute{Graduate Program in Applied Physics, Federal University of Latin-American Integration, Avenida Tarqu\'inio Joslin dos Santos, 1000 - Polo Universit\'ario, Foz do Igua\c{c}u/PR, Brazil\label{addr1}
          \and Departamento de Engenharias e Ci\^encias Exatas, Universidade Federal do Paran\'a, Rua Pioneiro, 2153, Palotina/PR, Brazil\label{addr2}
         }

\date{Received: date / Accepted: date}

\maketitle

\begin{abstract}
We investigate gravitational wave emission from an oscillating star modeled in a static Weyl background using a modulated waveform ansatz. First, we revisit the Weyl metric and obtain a Bessel-type curvature solution with regularity and asymptotic flatness conditions compatible for an astrophysical compact object. Second, we model an oscillating star-like object with a perfect fluid as a source of gravitational radiation. A modified wave equation with curvature-induced redshifts is solved numerically leading to a waveform that produces primary components due to wave trapping. This signal is injected into real LIGO data for GW150914 and compared across detectors allowing the detectability of curvature-induced gravitational waveform in the threshold of LIGO's sensitivity, consistent with an exotic compact object (ECO).

\end{abstract}

\section{Introduction}
The analysis of the sign received by Laser Interferometer Gravitational-wave Observatory~(LIGO) marked the very first direct evidence of the existence of gravitational waves~(GW) and black-holes~(BHs)~\cite{GWPRL2016}. With a very short signal duration around only 0.2 seconds, but with an extremely clear Signal-to-Noise Ratio (SNR), the sign originated from stellar-mass binary BH mergers confirmed one of the greatest predictions of Einstein's general relativity~(GR), opening a brand new gravitational wave astronomy era. Now, the observational efforts are focused on improving detector sensitivity for LIGO and Virgo~\cite{VIRGO2015}, and the development of future detectors like Laser Interferometer Space Antenna~ (LISA)~\cite{LISA2017} and Kamioka Gravitational Wave Detector~(KAGRA)~\cite{KAGRA2019}.  From the theoretical point of view, GW physics can produce deep questions/answers about the inner nature of gravity and compact objects. Thus, it is consolidating an arena for testing GR in strong-field regimes, as well as modified theories such as scalar-tensor and $f(R)$ gravity~\cite{Sotiriou2010,DeFelice2010}. Exotic compact object like boson stars and wormholes may also be explored via gravitational wave echoes and ringdown analysis~\cite{Cardoso2016} as well as deviations from the Kerr metric and potential horizonless objects~\cite{Mazur2004,Johannsen2011,Cardoso2016,Cardoso2019}. In GW context, quantum gravity may also be explored with concepts like soft hair, black hole remnants, and entanglement-based spacetime~\cite{Yunes2016,Barack2019}.

An alternative route to follow is to study the behavior of GWs in cylindrically symmetric spacetimes, particularly those described by Weyl metrics~\cite{Weyl1918,Stephani2003}, which may provide a rich framework for exploring exact solutions in GR and modified gravity theories~\cite{Kegeles1979, Gleiser1989, Nollert1999, Bronnikov2001,Stephani2003}. Physical implications may be achieved for, e.g., as the behavior of anisotropic gravitational and electromagnetic waves at very high energies~\cite{MestraPaez2023}, GW propagation within the Weyl invariant theory~\cite{Barros2024}, and cosmic anisotropy~\cite{Colgain2023}.

In this paper, we revisit the Weyl metric and propose new possible exact solutions exploring the possibilities of GW emission in such a space-time. In Section 2, we solve Einstein's equation obtaining the dynamical system of the Weyl potentials and the related effective potential using the Hamilton-Jacobi approach. In Sections 3 and 4, we propose new exact solutions to the Weyl metric, and analyze the causal structure to select a more appropriate solution to reproduce an astrophysical viable compact object. In the section 5, we study the GW production of the selected model obtaining the related Weyl-induced wave equation with a perfect fluid as a source. We develop a toy model of an oscillating star via numerical implementation. To this matter, we use \texttt{Python}-based codes such as \texttt{Bilby}~\cite{bilby} and \texttt{Dynesty}~\cite{Speagle_2020} sampler for Bayesian analysis and fitting. Moreover, we also use \texttt{Gwosc}~\cite{Abbott:2021:LIGO12,Abbott:2023:LIGO3} for database and \texttt{GWpy}~\cite{gwpy} for simulating time series data from LIGO Hanford (H1) and Livingston (L1) detectors. Finally, in the conclusion section we present the final remarks and prospects.

\section{Equations of motion and effective potential}
The non-rotating Weyl metric~\cite{Weyl1918} is given by:
\begin{equation}\label{eq:norot_weyl}
ds^2 = e^{2(\lambda - \sigma)} dr^2 + r^2 e^{-2\sigma} d\theta^2 + e^{2(\lambda - \sigma)} dz^2 - e^{2\sigma} dt^2\;.   
\end{equation}
This is a static, axially symmetric metric, which describes a static gravitational field, where \( \lambda(r, z) \) and \( \sigma(r, z) \) are the Weyl potentials in the cylindrical coordinates \( r \) and \(z \).  

Using Einstein's vacuum equations, one can find the exterior gravitational field given by the set of the simplified equations
\begin{align}
\hspace{3.5cm}    
\sigma_r + r(\sigma_{rr}+\sigma_{zz})&=0\;, \label{eq:01}\\
r(\sigma_r^2-\sigma_z^2)-\lambda_r&=0\;, \label{eq:02}\\
2r\sigma_r\sigma_z&=\lambda_z\;, \label{eq:03}
\end{align}
where we denote \((,r), (,z),(,rr),(,zz)\) the first and the second derivatives with respect to coordinates \((r,z)\), respectively. This set of equations was originally obtained by Weyl and investigated by Rosen~\cite{Rosen1949} in the problem of the motion and equilibrium of a particle in such a space-time. Analyzing the general form of these equations, we have that Eq.\eqref{eq:01} is just the Laplace equation in cylindrical coordinates for the Weyl potential \( \sigma(r, z) \), which behaves as a Newtonian-like gravitational potential. The other two equations, Eq.\eqref{eq:02} and Eq.\eqref{eq:03} are quadratures, which means that once \( \sigma(r, z) \) is known as a solution to the Laplace equation, \(\lambda(r,z)\) can be found by integrating these two equations. We point out that Eq.\eqref{eq:01} is consequence of Eqs.\eqref{eq:02} and \eqref{eq:03} for $\sigma_z(r,z)\neq 0$.

We start with a general determination of the equations of motion for the Weyl metric in Eq.\eqref{eq:norot_weyl}. By means of the Hamilton-Jacobi approach, we have that a motion of a test particle with mass \(\mu\) of a metric \(g_{\mu\nu}\) is given by
\begin{equation}\label{eq:metricHJ}
g^{\mu\nu} \frac{\partial S}{\partial x^\mu} \frac{\partial S}{\partial x^\nu} = -\mu^2\;.
\end{equation}
For separability, we assume the ansatz
\begin{equation}\label{eq:ansatz}
S = -E t + L \phi + P_z z + S_r(r) + S_z(z)\;,
\end{equation}
where \(E = -p_t\) is the energy (conserved due to time-translation symmetry), \(L = p_{\phi}\) is the angular momentum (conserved due to axial symmetry), \(P_z = p_z\) is the axial momentum (conserved due to \(z\)-translation symmetry), \(S_r(r)\) and \(S_z(z)\) are functions to be determined.

Substituting Eq.\eqref{eq:ansatz} into Eq.\eqref{eq:metricHJ}, we impose separability by assuming
\begin{align}\label{eq:ansatz2}
\hspace{0.6cm}
(\partial_r S_r)^2 + \frac{e^{2\lambda - 2\sigma}}{r^2} L^2 - e^{2\lambda - 4\sigma} E^2 + \mu^2 e^{2(\lambda - \sigma)}\nonumber\\ = -(\partial_z S_z)^2 - P_z^2.
\end{align}
The left side depends only on \(r\), while the right side depends only on \(z\). Thus, both must equal to a separation constant \(Q\) (the Carter constant) such as
\begin{align}\label{eq:ansatz3}
\hspace{1.5cm}
(\partial_r S_r)^2 = e^{2\lambda - 4\sigma} E^2 - \frac{e^{2\lambda - 2\sigma}}{r^2} L^2+ Q\nonumber \\- \mu^2 e^{2(\lambda - \sigma)} \;,\\
(\partial_z S_z)^2 = -P_z^2 - Q\;.
\end{align}

The four resulting integrals of motion are as follows. We define the energy \(E\) as the first integral
\begin{equation}\label{eq:energy}
E = e^{2\sigma} \dot{t},
\end{equation}
that confirms that energy \(E\) is conserved because the metric is stationary.  The dot symbol denotes a derivative with respect to an affine parameter $\tau$. The second integral of motion is given by the equation of motion for \(t\) that is
\begin{equation}\label{eq:time}
\dot{t} = E e^{-2\sigma}.
\end{equation}
Due to axial symmetry, we define the conserved angular momentum \(L\) and write the relation
\begin{equation}\label{eq:phi}
\dot{\phi} = \frac{L e^{2\sigma}}{r^2}\;.
\end{equation}
And the final integral of motion is defined by the axial momentum \(P_z=e^{2(\lambda - \sigma)} \dot{z}\). Since momentum along \(z\) is conserved because the metric is translationally symmetric in the coordinate \(z\), we obtain the equation of motion for \(z\) as
\begin{equation}\label{eq:zequation}
\dot{z} = P_z e^{-2(\lambda - \sigma)}\;.
\end{equation}
Then, equations Eq.\eqref{eq:energy},\eqref{eq:time},\eqref{eq:phi} and \eqref{eq:zequation} describe the motion of a test particle in the Weyl metric in terms of the conserved quantities \((E, L)\). Particularly, the time evolution equation in Eq.\eqref{eq:time} shows that time is affected by the gravitational potential \( \sigma(r,z)\), which means that a particle closer to strong gravitational sources and clocks run slower due to gravitational time dilation. On the other hand, far from the source (large \( r \), large \( z \)), \( e^{2\sigma} \to 1 \), and time evolution asymptotically approaches that of flat Minkowski space.

From separability, we also obtain the equation of motion for radial coordinate \(r\) as
\begin{align}\label{eq:radialeq}
\hspace{0.5cm}  \dot{r}^2 = e^{-2(\lambda - \sigma)} \left( Q + E^2 e^{-2\sigma} - \frac{L^2}{r^2 e^{2\sigma}} - \mu^2\right) \nonumber \\ 
                               + e^{-4(\lambda - \sigma)}P_z^2\;.
\end{align}
The Carter constant \(Q\) can be seen as a measure of non-planar motion (deviation from purely equatorial geodesics). In other words, it ensures that the radial and vertical \(z\)-motions decouple. 

To interpret Eq.\eqref{eq:radialeq} as an energy conservation equation, we compare it to the Newtonian form
\begin{equation}\label{eq:newtform}
\dot{r}^2 + V_{\text{eff}}(r,z) = \mathcal{E},
\end{equation}
where $\mathcal{E}$ is the effective energy and \(V_{\text{eff}}(r,z)\) is the effective potential. Thus, from the radial equation of Eq.\eqref{eq:radialeq}, we identify the effective potential \(V_{\text{eff}}(r,z)\) as
\begin{align}\label{eq:effpot}
V_{\text{eff}}(r,z) = e^{-2(\lambda-\sigma)} \left( \frac{L^2}{r^2 e^{2\sigma}} + \mu^2 - Q - E^2 e^{-2\sigma} \right)\nonumber \\
                          -  e^{-4(\lambda - \sigma)}P_z^2\;,
\end{align}
and \(\mathcal{E} = 0\) due to the constraint \(p_\mu p^\mu = -\mu^2\). Stable orbits occur where \(V_{\text{eff}}(r,z)\) has a minimum, and turning points (where \(\dot{r} = 0\)) are given by \(V_{\text{eff}}(r,z) = 0\). The Carter constant \(Q\) acts as a repulsive potential for \(Q > 0\) and attractive for \(Q < 0\), though \(Q \geq 0\) for physical motion.

We obtain some special cases, such as the planar motion (\(P_z = 0\), \(Q = 0\)). If the particle moves only in the \((r, \phi)\) plane (no \(z\)-motion), the potential simplifies to
\begin{equation}\label{eq:veff2}
V_{\text{eff}}(r) = e^{2(\sigma - \lambda)} \left( \frac{L^2}{r^2 e^{2\sigma}} + \mu^2 - E^2 e^{-2\sigma} \right)\;,
\end{equation}
while the radial free fall occurs at (\(L = 0\), \(P_z = 0\), \(Q = 0\)) and we have
\begin{equation}\label{eq:veff3}
V_{\text{eff}}(r) = e^{2(\sigma - \lambda)} \left( \mu^2 - E^2 e^{-2\sigma} \right).
\end{equation}

Concerning horizons, they occur where the component \( g_{tt} = 0 \), meaning that \(e^{2\sigma} = 0\). Since \( e^{2\sigma} \) is always positive for finite values of \( r \) and \( z \), and \textit{unless a different choice of potentials} \(\sigma(r,z)\) and \(\lambda(r,z)\) \textit{is made}, there are no horizons in this spacetime. This suggests that the solution may describe a naked singularity rather than a black hole. In the Schwarzschild metric, it does not admit naked singularities, as the singularity is hidden behind the event horizon. On the other hand, Curzon-Chazy~\cite{Chazy1924,Curzon1925} and the Zipoy-Voorhees~\cite{Zipoy1966,Voorhees1970} metrics generally introduce naked singularities in their standard forms. Next, we present two different methodologies to solve the system of the field equations~\eqref{eq:01}, \eqref{eq:02} and \eqref{eq:03}. Then, we analyze the behavior and consequences of different Weyl potentials \(\sigma(r,z)\) and \(\lambda(r,z)\).

\section{Radial solution}  
Inspired in the group-invariant solution theory (see chapter 3 in Ref.\cite{MR836734}), which reduces PDE to ODE, we present the first methodology to directly solve the system of equations~\eqref{eq:01}, \eqref{eq:02} and \eqref{eq:03}. Assuming that $\sigma(r,z) = \tilde{\sigma}({r^2+z^2})$, we introduce the change of variable $\bar{t}=r^2+z^2$. By means of the chain rule, we obtain,
\begin{align}
\hspace{3cm}   
\sigma_r= \frac{d \tilde{\sigma}}{d \bar{t}}\frac{\partial  (r^2+z^2)}{\partial r}=2r\tilde{\sigma}'\;, \label{eq:sigmar}\\ 
\sigma_{rr}= 2\tilde{\sigma}' + 4r^2\tilde{\sigma}''\;,\label{eq:sigmarr}\\
\sigma_{zz} = 2\tilde{\sigma}' + 4z^2\tilde{\sigma}''\;,\label{eq:sigmazz}
\end{align}
where prime and double-prime denote the first and second derivatives of the real function $\tilde{\sigma}(\bar{t})$, respectively. Substituting Eqs.\eqref{eq:sigmar},\eqref{eq:sigmarr} and \eqref{eq:sigmazz} in Eq.\eqref{eq:01}, we have
\begin{equation}
6\tilde{\sigma}'(r^2+z^2) + 4 (r^2 +z^2)\tilde{\sigma}''(r^2+z^2)=0\;.    
\end{equation}
Then, Eq.\eqref{eq:01} is reduced to the ODE
\begin{equation}\label{eq:ode1}
3\tilde{\sigma}'(\bar{t}) + 2 t\tilde{\sigma}''(\bar{t}) = 0 . 
\end{equation}
that gives immediately
\begin{equation}\label{eq:sigmaradial}
\sigma(r,z) = \frac{k_1}{\sqrt{r^2+z^2}} + k_2, 
\end{equation}
where \(k_1,  k_2\) are integration constants with \(k_1,  k_2\in \mathbb{R}\).

From Eq.\eqref{eq:03}, there exist a function $h(r)$ such that
\begin{equation}\label{eq:lambdaradial}
\lambda(r,z) = -\frac{k_1^2r^2}{2(r^2+z^2)^2}+ h(r)\;.
\end{equation}
Using Eq.\eqref{eq:02}, we obtain that $h(r)=k_3\in \mathbb{R}$ in which \(k_3\) is an integration constant.

Interestingly, the given Weyl potentials \(\sigma(r, z)\) and \(\lambda(r, z)\) generalizes very known solutions as Schwarzschild and Curzon-Chazy solutions. The potential \(\sigma(r, z)\) suggests a monopole-type solution in Weyl coordinates, analogous to the Newtonian potential of a point mass. The first term of $\sigma$ in  Eq.\eqref{eq:sigmaradial} resembles the potential of a Schwarzschild-like object in these coordinates. The constant \( k_2 \) shifts the potential but does not introduce necessarily new physics. On the other hand, \(\lambda(r, z)\) modifies the Weyl metric's conformal factor and is often required to satisfy Einstein's equations. It also generalizes the Curzon-Chazy metric with the appearance of the \(k_3\) integration constant. 

\subsection{Causal structure}

A more physically appealing form can be found when we replace the integration constants in Eq.\eqref{eq:sigmaradial} and Eq.\eqref{eq:lambdaradial} by physical constants correlated with mass \( M \), such as \( k_1= M \) and keeping \( k_2, k_3 \) as dimensionless constants. Therefore, the form of the potentials \( \sigma(r, z) \) and \( \lambda(r, z) \) becomes
\begin{align}
\hspace{2.75cm}\sigma(r, z) = \frac{M}{\sqrt{r^2 + z^2}}+ k_2\;,\label{eq:sigmarad}\\
\lambda(r, z) = -\frac{M^2 r^2}{2 (r^2 + z^2)^2} + k_3\;.\label{eq:lambdarad}
\end{align}
As pointed out before, since \(e^{2\sigma(r,z)}\) is always positive, this means that the Weyl metric does not have an event horizon unless a different choice of potentials is made. Here, the introduction of \( k_2 \) shifts the potential at large distances (as \( r \to \infty \)) and determines the asymptotic flatness or gravitational potential at large distances. 

The horizons are given by the null surface condition \(g_{tt} = -e^{2\sigma(r, z) } = 0\). This implies that \(e^{2\sigma(r, z) } = 0\) leads to \(\sigma(r, z)  \to -\infty\). For a positive mass \(M > 0\), \(\sigma(r, z)  \to -\infty\) only if \(\sqrt{r^2 + z^2} \to 0\) (i.e., \(r \to 0\) and \(z \to 0\)). However, \(\sigma(r, z) \) diverges to \(+\infty\) as \(r, z \to 0\), then no horizon exists unless \( k_2 \) is negative. For the case \( k_2 = 0\), \(\sigma(r, z) \) is always finite and positive for any \((r, z) \neq 0\). At the origin \(r = z = 0\), \(\sigma(r, z)  \to +\infty\), meaning that \(g_{tt} \to -\infty\) is a singularity. When \( k_2 < 0\), if we consider \( k_2 = -| k_2 |\), then
\begin{equation}
  \sigma(r,z) = \frac{M}{\sqrt{r^2 + z^2}} - | k_2 |.  
\end{equation}
and \(\sigma(r, z)  \to -\infty\) when \(\frac{M}{\sqrt{r^2 + z^2}} = | k_2 |\), i.e., at
\begin{equation}
  \sqrt{r^2 + z^2} = \frac{M}{| k_2 |}.
\end{equation}
This defines a spherical horizon at radius \(R_H = \frac{M}{| k_2 |}\).
The location of the horizon is at \(r^2 + z^2 = \left(\frac{M}{| k_2 |}\right)^2\), which is a sphere in \((r, z)\) coordinates. To check if it is a black hole horizon, the surface gravity \(\kappa\) is
\begin{equation}
  \kappa = \sqrt{-\frac{1}{2} (\nabla_\mu \xi_\nu)(\nabla^\mu \xi^\nu)},
\end{equation}
where \(\xi = \partial_t\) is the timelike Killing vector. On the horizon, \(g_{tt} = 0\), so \(\xi\) becomes null. For our metric, \(\kappa\) is non-zero, confirming a non-degenerate horizon. 

On the other hand, the \(\lambda(r, z) \) potential modifies the spatial geometry but does not affect the horizon location. However, it influences the proper distance to the horizon and the geometry of spatial slices. For \( k_2 < 0\), the spacetime has a horizon at \(R_H = \frac{M}{| k_2 |}\), resembling an asymptotically anti-de Sitter (AdS) black hole. The singularity at \(r = z = 0\) is hidden behind the horizon. For \( k_2 = 0\), no horizon exists, and the singularity at \(r = z = 0\) is naked. For \( k_2 > 0\), the potential \(\sigma(r, z) \) remains finite everywhere, and no horizon forms.

To determine the presence of curvature singularities, we calculate the Kretschmann scalar \( K = R_{\mu\nu\rho\sigma} R^{\mu\nu\rho\sigma} \), where \( R_{\mu\nu\rho\sigma} \) is the Riemann tensor. We find that the leading-order divergence is given by
\begin{equation}
   K \approx \frac{C M^2}{(r^2 + z^2)^3} + \text{subleading terms},
\end{equation}
where \( C \) is a numerical constant. Hence, the Kretschmann scalar diverges as \( r, z \to 0 \) similar to the Schwarzschild metric \( K \sim M^2/r^6 \) at \( r \to 0 \). Along the z-axis, the singularity is string-like at \( r = 0 \).
\begin{figure*}[t!]
\centering
\includegraphics[width=6.8in, height=2.6in]{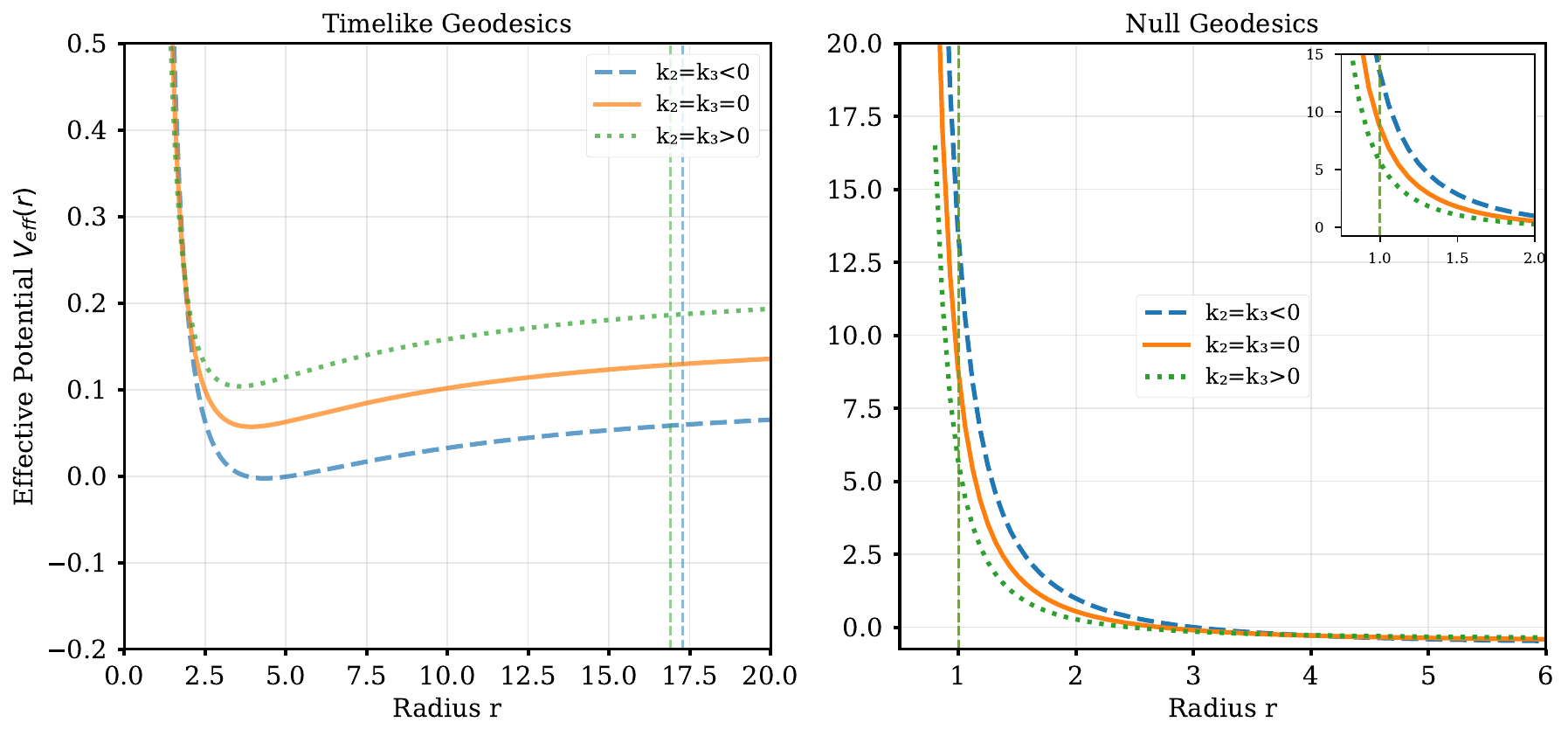}
\caption{Effective potential for \(E\) < \(L\) of dimensionless constants when $k_2 = k_3$ for massive particles (left panel) and photons (right panel). In the timelike geodesics, vertical dashed lines indicate ISCO. In the null geodesics panel, the photon sphere radius is $r_{ph}=1$ as represented by the green dashed vertical line.} \label{fig:effpot1}
\end{figure*}

This confirms our previous results, which show that there is a curvature singularity at the origin \( (r, z) = (0, 0) \). For \( k_2 < 0 \), this singularity is hidden behind a horizon, and for \( k_2 \geq 0 \), the singularity is naked. However, if \( k_2 \neq k_3 \), the spatial curvature deviates from the causal structure, modifying geodesic properties and potentially leading to anisotropies in curvature evolution. 

In Fig.(\ref{fig:effpot1}), we present the results for effective potential as a function of the radius for massive particles and photons. In the analysis of different gravity regimes, the behavior of both timelike and null geodesics is explored across three distinct cases: strong gravity ($k_2 = k_3$ < 0), Schwarzschild-like ($k_2 = k_3$ = 0), and weak gravity ($k_2 = k_3$ > 0). For the strong gravity regime, it presents an unusual large ISCO, 
more commonly expected for retrograde orbits around high-spin BHs, modified gravity theories or exotic compact objects. The ISCO radius is found to be at 17.29 with a corresponding angular momentum of 3.58, while the photon sphere occurs at \(r=1\) with angular momentum of 3. Timelike geodesics show that stable circular orbits are only possible for radii greater than 17.29, with plunging orbits observed at smaller radii. In the photon case, photons are captured for radii below 1 and deflected above it, with the unstable circular photon orbits located at \(r = 1\). 

In the Schwarzschild-like case, no ISCO was found, but the photon sphere remains at \(r = 1\) with an angular momentum of 2.72. Photon orbits in this case behave similarly, with photons being captured below \(r = 1\) and scattered above. In the weak gravity regime, the ISCO radius decreases to 16.90 with angular momentum of 4.26, and the photon sphere occurs at \(r = 1\) with angular momentum of 2.46. Similar to the strong gravity case, stable circular orbits are found for radii above 16.90, while plunging orbits dominate at smaller radii. The photon behavior mirrors that of the previous two regimes, with photons being captured below 1, unstable at \(r=1\), and scattered above that value. While the weak gravity scenario might present an astrophysical viable configuration, both strong and Schwarzschild-like regimes are ruled out in explaining compact object environments observed via electromagnetic and GW signatures. 

\begin{figure*}[ht!]
\centering
\includegraphics[width=6.8in, height=2.6in]{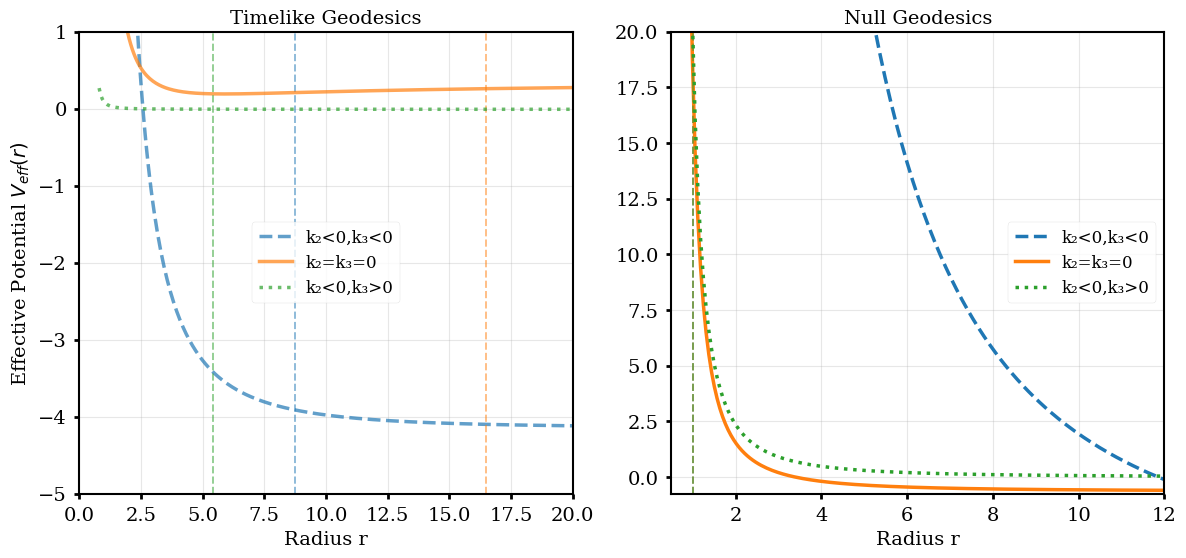}
\caption{Effective potential for \(E\) < \(L\) of dimensionless constants when $k_2 \neq k_3$ for massive particles (left panel) and photons (right panel). In the timelike geodesics, vertical dashed lines indicate ISCO. In the null geodesics panel, the photon sphere radius is $r_{ph}=1$ as represented by the green dashed vertical line. The $k_2=k_3$ case of orange line is included for comparison.} \label{fig:effpot2}
\end{figure*}

In Fig.(\ref{fig:effpot2}), as in the cases before, we present the results for effective potential as a function of the radius for massive particles and photons for \(k_2 \neq k_3\). In our findings, we explore two viable cases for analyzing when \(k_2\) and \(k_3\) are negatives (but with different values), and with \(k_2<0\) and \(k_3>0\). For the case \(k_2>0\) and \(k_3<0)\) we have a very large nonphysical ISCO, and we omitted it accordingly. We maintain the ``Schwarzschild'' case ($k_2 = k_3$ = 0), just for comparison purposes.

For \(k_2<0, k_3<0\) (dashed blue line), in the timelike case (left panel), there is a minimum around \( r \approx 8.76 \). It presents a deep potential well but supports stable orbits beyond ISCO and gradually becomes Newtonian-like at large \( r \). For null geodesics, \(V_{\text{eff}} \) shows a peak indicating unstable photon orbits at \( r = 1 \) with manageable photon angular momentum \( L_\text{photon} \approx 9.49 \). On the other hand, the case \( k_2<0, k_3>0 \) (dotted green line) presents a steep drop and sharp minimum for \(V_{\text{eff}} \) of massive particle at small radius \( r \approx 5.43 \), suggesting a strong gravitational field, which might lead to extreme effects and high instability near the core. For photon potential, it peaks at \( r = 1 \) but presents an extreme sharp spike and a huge \( L_\text{photon} = 27.11 \) suggesting a very strong deflection, that may reveal possible photon trapping or chaos near the center.

All the cases analyzed so far, the radial solution for Weyl potentials indicate the preferred values for \(k_2<0, k_3<0\) (\(k_2\neq k_3\)). It produces a system that may support realistic disk formation and shadow structure. The ISCO is reasonably balanced enough for efficient accretion, but not unnaturally small or large.  At \( r=1 \), the photon aligns with GR predictions. An important aspect that the moderate value of \( L_\text{photon} \) avoids extreme gravitational lensing as well. 

\section{Oscillatory and exponential solutions}
In order to find a more regular to model a possible compact object, we propose an alternative solution of Eq.~\eqref{eq:01}, and taking into account the symmetry structure of this spacetime, we apply the classic technique of separation of variables in such we have the ansatz 
\begin{equation}\label{eq:sigmaseparation}
\hspace{3cm}  \sigma(r, z) = f(r)g(z)\;.
\end{equation}
Substituting this form into Eq.~\eqref{eq:01} and assuming \( r \neq 0 \), we rearrange the terms and find that both sides of the equation depend on different variables, separated with the constant \( K \in \mathbb{R} \). This gives us the relation
\begin{equation}
\hspace{3cm}   - \frac{f_r + r f_{rr}}{r f} = \frac{g_{zz}}{g} = K\;,
\end{equation}
where $f=f(r)$ and $g=g(z)$. From this, we get two separate ordinary differential equations 
\begin{align}
\hspace{3cm} r f_{rr} + f_r + K r f &= 0, \label{eq:f(r)} \\
    g_{zz} - K g &= 0\;. \label{eq:g(z)}
\end{align}
These equations are much simpler to handle than the original PDE, and depending on the sign of \( K \), we end up with interesting case solutions.

\subsection{\bf Case $K=0$} 
Solving the system of Eq.\eqref{eq:f(r)} and \eqref{eq:g(z)}, there exist real integration constants $a, b, c$ and $d$ such that 
\begin{align}
    g(z )&=a z + b,\\
    f(r)&= c \ln r + d\;.
\end{align}
Then, we have immediately the \(\sigma(r, z)\) potential 
\begin{equation}\label{eq:sigmasep00}
\sigma(r, z) = (a_0 + a_1 z) \ln r + a_2 z+ a_3 \;,
\end{equation}
where, for simplicity, we denote the constants \(a_0=c.b\), \(a_1=a.c\), \(a_2=a.d\) and \(a_3=b.d\). Moreover, from Eq.\eqref{eq:02}, we obtain the \(\lambda(r, z)\) potential as
\begin{align}
\lambda(r,z)&= (c \ln r +d)^2\frac{r^2}{2} -c(c\ln r + d)\frac{r^2}{2} +  c^2\frac{r^2}{4} + h(z),
\end{align}
where $h(z)$ is obtained from Eq.\eqref{eq:03}, and denoted as \(h(z)=cd(az+b)^2 +l,\quad  l\in \mathbb{R}\). Since logarithm potentials in Eq.\eqref{eq:sigmasep00} have been extensively explored in literature~\cite{LyndenBell1962,Mestel1963,Ostriker1964,Shakura1973,Milgrom1983a,Milgrom1983b,Milgrom1983c,BekensteinMilgrom1984,Evans1993,Navarro1996,Frank2002,Binney2008,CapistranoBarrocas2018,Capistrano2019,CapistranoKalbCoimbraAraujo2022}, in the following, we focus only on the other two new cases.

\subsection{\bf Case $K>0$}  
One can immediately integrate Eq.\eqref{eq:g(z)}, and obtain the hyperbolic solution
\begin{equation}\label{eq:sigmasep01}
g(z)=A e^{\sqrt{K}z} + Be^{-\sqrt{K}z}\;.
\end{equation}
Using a substitution $f(r)=\tilde{f}(\sqrt{K}r)$,  Eq.\eqref{eq:f(r)} is equivalent to the zeroth-order Bessel equation. Taking the variable change $t=\sqrt{K}r$, we have
\begin{align}\label{eq:bessel}
    t^2\tilde{f}_{tt} + t\tilde{f}_t + t^2\tilde{f}=0\;.
\end{align}
For $r>0$, we have the general solution 
\begin{align}\label{eq:bessel2}
\tilde{f}(t)=C_1J_0(t) + C_2Y_0(t)\;,
\end{align}
where $(C_1,C_2)$ are integration constants, and 
\begin{align}
J_0(t)&=1+\sum_{m=1}^{\infty}\frac{(-1)^mt^{2m}}{2^{2m}(m!)^2},\\
    Y_0(t)&=\frac{2}{\pi}\left[\left(\gamma + \ln \frac{t}{2}\right)J_0(t) + \sum_{m=1}^{\infty}\frac{(-1)^{m+1}H_m}{2^{2m}(m!)^2}t^{2m}\right],
\end{align}
where the functions $J_0(t)$ and $Y_0(t)$ are the zeroth-order Bessel functions of the first and second kind, respectively. The term $\gamma$ is the Euler-Máscheroni constant and $H_m$ represents the $m$-th harmonic number given by $H_m=1+\frac{1}{2}+\frac{1}{3} + \ldots + \frac{1}{m}$. Thus, for $K>0$, the \(\sigma(r, z)\) potential can be written generically as 
\begin{equation}\label{eq:sigmasep01}
\sigma(r, z) = (A e^{\sqrt{K}z} + Be^{-\sqrt{K}z})(C_1J_0(\sqrt{K}r) + C_2Y_0(\sqrt{K}r)) \;.
\end{equation}
and consequently from Eq. \eqref{eq:02} and Eq. \eqref{eq:03},
\begin{align}
\lambda(r,z)&=r\sqrt{K}\zeta_0(r)\zeta_1(r)(A e^{\sqrt{K}z} + Be^{-\sqrt{K}z})^2\nonumber\\&+4ABK\left[\frac{1}{2}C_1^2r^2 \zeta_{01}(r)\right.+ C_1C_2r^2\varsigma_{0011}(r)\nonumber\\& \left.+ \frac{1}{2}C_2^2r^2 \varsigma_{01}(r)\right] 
\end{align}
where, just for the sake of notation, we denote the functions
\begin{eqnarray}
\zeta_0 (r)=C_1J_0(\sqrt{K}r) + C_2Y_0(\sqrt{K}r), \\
\zeta_1(r) = -C_1J_1(\sqrt{K}r)- C_2Y_1(\sqrt{K}r), \\
\zeta_{01}(r)= J_0(\sqrt{K}r)^2+J_1(\sqrt{K}r)^2,\\
\varsigma_{0011}(r)= J_0(\sqrt{K}r)Y_0(\sqrt{K}r) + J_1(\sqrt{K}r)Y_1(\sqrt{K}r), \\
\varsigma_{01}(r)=Y_0(\sqrt{K}r)^2+Y_1(\sqrt{K}r)^2. 
\end{eqnarray}

We start to study the causal structure of the solution \( K > 0 \) by analyzing the boundary conditions and regularity at \( r = 0 \). The behavior at origin \( r \to 0 \), we have that \(J_0(\sqrt{K} r) \approx 1 - \frac{K r^2}{4} + \mathcal{O}(r^4)\) is regular. On the other hand, \(Y_0(\sqrt{K} r) \approx \frac{2}{\pi} \left( \gamma + \ln \left( \frac{\sqrt{K} r}{2} \right) \right) + \mathcal{O}(r^2 \ln r)\) diverges logarithmically. Thus, for regularity at \( r = 0 \), we must set \( C_2 = 0 \), and Eq.\eqref{eq:sigmasep01} is simplified as
\begin{equation}\label{eq:sigmasep02}
\sigma(r,z) = C_1 J_0(\sqrt{K} r) \left( C \cosh(\sqrt{K} z) + D \sinh(\sqrt{K} z) \right) \;.
\end{equation}
This is smooth at \( r = 0 \), since \( J_0 \to 1 \), and the exponential part depends only on \( z \). Similarly, the divergences in \( \lambda \) vanish when \( C_2 = 0 \), simplifying to a combination of finite (\( J_0 \), \( J_1 \)) terms. The asymptotic behavior (\( z \to \infty \), \( r \to \infty \)) leads to \( \sigma \to 0 \) and \( \lambda \to 0 \) at large distances. If \( A \neq 0 \), \( \sigma \sim e^{\sqrt{K}z} \to \infty \) when \( z \to \infty \). Thus, for asymptotic decay in \( z \), we set \( A = 0 \). Then,
\begin{align}\label{eq:sigmabessel}
\sigma(r, z) = \alpha_0 e^{-\sqrt{K}z} J_0(\sqrt{K}r)\;,
\end{align}
where we denote $\alpha_0=B C_1$ as a parameter from integration constants $B$ and $C_1$. We point out that the solution in Eq.\eqref{eq:sigmabessel} vanishes exponentially as \( z \to \infty \), mimicking asymptotic flatness in the vertical direction. For fixed \(z\), when \( r \to \infty \), Bessel functions decay as
\begin{align}
  J_0(\sqrt{K}r) \sim \sqrt{\frac{2}{\pi \sqrt{K}r}} \cos\left(\sqrt{K}r - \frac{\pi}{4}\right)
\end{align}
So the potential decays with oscillations in a form \( 1/\sqrt{r} \). Hence, \( \sigma \to 0 \) decays, but not monotonically. Thus, we get a quasi-flat behavior with oscillations similar to cylindrical wave tails.

For \( \lambda(r, z) \) potential, we start by imposing regularity at \( r = 0 \). To eliminate singularities at the axis, we set \( C_2 = 0 \) and remove the functions (\( Y_0, Y_1 \)), which diverge at \( r = 0 \). Then, the expression becomes
\begin{align}
\lambda(r, z) &= -r C_1^2 J_0(\sqrt{K} r) J_1(\sqrt{K} r) (A e^{\sqrt{K} z} + B e^{-\sqrt{K} z})^2 \nonumber\\
&\quad + 4ABK r^2 \left[ \frac{1}{2} A^2 \left( J_0^2(\sqrt{K} r) + J_1^2(\sqrt{K} r) \right) \right]
\end{align}

To achieve asymptotic decay, we set \( A = 0 \) to eliminate growing exponential term \( e^{\sqrt{K} z} \). Thus, we have the simplified final expression
\begin{align}\label{eq:lambdabessel}
\lambda(r, z) = - \alpha_0^2 \, r \, J_0(\sqrt{K} r) J_1(\sqrt{K} r) \, e^{-2\sqrt{K} z}\;.
\end{align}

\subsection{\bf Case $K<0$} 
In a similar procedure, from Eq.\eqref{eq:g(z)}, one has
\begin{equation}\label{eq:sigmasep05}
g(z)=A \cos{\sqrt{-K}z} + B\sin{\sqrt{-K}z}\;.
\end{equation}
On the other hand, Eq. \eqref{eq:f(r)} becomes a modified Bessel equation of order $0$ (see section 6.4 in Ref.\cite{zill2012first}) whose solutions are in the form
\begin{align}\label{eq:secondsolf(r)}{f}(r)=C_1 I_0(\sqrt{-K}r) + C_2 K_0(\sqrt{-K}r),\end{align}
where \(I_0(r)=J_0(ir)\). The Bessel function \(J_0(i \kappa r)\) is related to the modified Bessel function \(I_0(\kappa r)\), which grows exponentially with \(r\) as \(J_0(i \kappa r) = I_0(\kappa r)\). The function \(K_0(r)=\int_0^{\infty}\frac{\cos r \tau}{\sqrt{\tau^2+1}}d\tau,\) is the modified Bessel function of second kind of zeroth order. Hence, $\sigma(r,z)$ becomes
\begin{align}\sigma(r,z)&=\left(A \cos{\sqrt{-K}z} + B\sin{\sqrt{-K}z}\right)\nonumber\\
&\qquad\times \left(C_1 I_0(\sqrt{-K}r) + C_2 K_0(\sqrt{-K}r)\right),\end{align}
and consequently,
\begin{align}\label{lambda60}\lambda(r,z)=rff_rg^2 + K(A^2+B^2)\int rf^2dr,\end{align}
where $g=g(z)$ and $f=f(r)$ are given by Eq. \eqref{eq:sigmasep05} and \eqref{eq:secondsolf(r)} respectively. Note that after substitution $f(r)=\tilde{f}(i\sqrt{-K}r)$, Eq.\eqref{eq:f(r)} is equivalent to the $0$-order Bessel equation in Eq.\eqref{eq:bessel}. Hence, one gets the \(\sigma(r, z)\) potential in a form 
\begin{equation}\label{eq:sigmasep06}
\sigma(r, z) = \xi_1(z)\chi_1(r) \;,
\end{equation}
where we denote \(\xi_1(z)=\left(A \cos{\sqrt{-K}z} + B \sin{\sqrt{-K}z}\right)\) and \(\chi_1(r)=\left(C_1 I_0(\sqrt{-K}r) + C_2 K_0(\sqrt{-K}r)\right)\). Moreover, given the potential \(\sigma(r, z)\), and after a tedious calculation, one finds \(\lambda(r, z)\) in a form 
\begin{align}
\lambda(r, z) &= \frac{r^2 K}{2} \left[
 \xi_2^2(z) \chi_1^2(r)-\xi_1^2(z) \chi_2^2(r)\right]\nonumber\\
 &+\displaystyle\int r^2\chi_1^\prime(r)((\xi_1^\prime(z))^2\chi_1(r)-\xi_1^2(z)\chi_1^{\prime\prime}(r))dr,
 \end{align}
where we denote \(\xi_2(z)=\left(-A \sin{\sqrt{-K}z} + B \cos{\sqrt{-K}z}\right)\) and \(\chi_2(r)=\left(C_1 I_1(\sqrt{-K}r)-C_2 K_1(\sqrt{-K}r)\right)\). 

In order to have a glimpse of a physical reality, the potentials \(\sigma(r, z)\) and \(\lambda(r, z)\) must obey physical constraints when we analyze the causal structure, e.g., regularity in \(r = 0\) and decay/growth conditions in \(z\). The regularity at \(r = 0\) must not have divergence. Then, we start by setting \(C_2 = 0\) to eliminate \(Y_0(\sqrt{K}r)\), since \(Y_0\) diverges at \(r = 0\). The potential becomes
\begin{equation}
\sigma(r, z) = \left(A \cos{\kappa z} + B \sin{\kappa z}\right) C_1 J_0(i \kappa r)\;,
\end{equation}
where \(\kappa = \sqrt{-K} > 0\), since \(K < 0\). Thus, we have the form
\begin{equation}
   \sigma(r, z) = \left(A \cos{\kappa z} + B \sin{\kappa z}\right) I_0(\kappa r).
\end{equation}
Analyzing the causal structure, we need to check regularity on the origin \( r = 0 \) and for all \( z \in \mathbb{R} \). The behavior of the modified Bessel functions at \( r \to 0 \) shows that \( I_0(\kappa r) \to 1 + \frac{(\kappa r)^2}{4} + \dots \) is regular. On the other hand, \( K_0(\kappa r) \to -\log(\kappa r) + \text{const} + \dots \) has a logarithmic divergence. We obtain a similar situation for \( I_1(\kappa r)\) that is regular, and for \( K_1(\kappa r)\) is also divergent. Then, near \( r \to 0 \), Eq.\eqref{eq:secondsolf(r)} is regular only if \( C_2 = 0 \). Similarly, we analyze the derivative 
\begin{equation}
f_r(r) = \kappa \left[C_1 I_1(\kappa r) - C_2 K_1(\kappa r)\right]
\end{equation}
that diverges at \( r = 0 \), unless one sets \( C_2 = 0 \). Thus, for regularity at the axis, we must set \( C_2 = 0 \). 

Imposing the asymptotic flatness conditions (\( r \to \infty \), \( z \to \infty \)), one has the Bessel functions for large \( r \) as \( I_0(\alpha r) \sim \frac{e^{\alpha r}}{\sqrt{2\pi \alpha r}} \) that show an exponential growth, and for \( K_0(\alpha r) \sim \sqrt{\frac{\pi}{2\alpha r}} e^{-\alpha r} \), we obtain an exponential decay. Thus, to satisfy asymptotic flatness conditions, we must eliminate the growing \( I_0 \) term setting \( C_1 = 0 \). To
satisfy both regularity and asymptotic flatness, the only possible solution is the trivial one:
\begin{equation}
f(r) = 0 \quad \Rightarrow \quad \sigma(r,z) = 0, \quad \lambda(r,z) = 0\;.
\end{equation}
In summary, the solution $K<0$ presents an nonphysical behavior for infinite domain.

\section{Waveform injection and GW simulation}
\subsection{Oscillating star modeling}
In order to get insight on possible GW signs, we work in the linearized gravity limit. We derive the full gravitational wave equation for perturbations \( \bar{h}_{\mu\nu} \) in the Weyl metric, as such
\begin{equation}
g_{\mu\nu} = g^{(0)}_{\mu\nu} + h_{\mu\nu}, \quad \bar{h}_{\mu\nu} = h_{\mu\nu} - \frac{1}{2} g^{(0)}_{\mu\nu} h\;.
\end{equation}
Since a static vacuum solution does not produce GWs, we propose a model writing the linearized Einstein equation with a source as
\begin{equation}
\Box \bar{h}_{\mu\nu} + 2 R^{\lambda}{}_{\mu\nu}{}^{\rho} \bar{h}_{\lambda\rho} = -16\pi G \, \delta T_{\mu\nu}
\end{equation}
where \( \Box \) denotes the wave operator, \( R^{\lambda}{}_{\mu\nu}{}^{\rho} \) is the Riemann tensor, and \(\delta T_{\mu\nu}\) is the perturbed energy-momentum tensor. Thus, for each component \( \bar{h}_{\mu\nu} \), we have the generic wave equation for Weyl metric as
\begin{align}
e^{-2\sigma} \partial_t^2 \bar{h}_{\mu\nu}
- e^{-2(\lambda - \sigma)} \left( \partial_r^2 + \frac{1}{r} \partial_r + \partial_z^2 \right) \bar{h}_{\mu\nu}
\nonumber\\+ 2 R^{\lambda}{}_{\mu\nu}{}^{\rho} \bar{h}_{\lambda\rho}
= 16\pi G \, \delta T_{\mu\nu}
\end{align}
Or more explicitly for the indices \( \mu = \nu = t \), we have
\begin{align}\label{eq:gw02}
e^{-2\sigma} \partial_t^2 \bar{h}_{tt}
- e^{-2(\lambda - \sigma)} \left( \partial_r^2 + \frac{1}{r} \partial_r + \partial_z^2 \right) \bar{h}_{tt}
\nonumber\\- 2 e^{-2(\lambda - \sigma)} \left[ \left( \sigma_{,rr} + \frac{1}{r} \sigma_{,r} \right) \bar{h}_{rr} + \sigma_{,zz} \bar{h}_{zz} \right]
= 16\pi G \, \delta T_{tt}
\end{align}
Clearly, the wave operator is redshifted by the background Weyl's potentials \( \sigma(r,z)\) and \(\lambda(r,z)\). To complete Eq.\eqref{eq:gw02}, we need the explicit form of \( \delta T_{tt} \). Thus, we assume a perfect fluid on a perturbed background.
The full stress-energy tensor is
\begin{align}\label{eq:energytensor}
T_{\mu\nu} = (\rho + p) u_\mu u_\nu + p g_{\mu\nu}
\end{align}
where \( \rho \) is energy density, \( p \) is the  pressure and \( u^\mu \) is the four-velocity of the fluid (normalized as \( u^\mu u_\mu = -1 \)). 

Following the standard procedure, we have the perturbation \( \delta T_{\mu\nu} \) \(T_{\mu\nu} = T^{(0)}_{\mu\nu} + \delta T_{\mu\nu}\) produced from the perturbation in the metric \( \delta g_{\mu\nu}\) and the and the perturbation in the fluid variables \( \delta \rho, \delta p, \delta u^\mu \). It leads to the general form for \( \delta T_{tt} \)
\begin{align}\label{eq:tmunu01}
\delta T_{tt} = e^{2\sigma} \delta \rho - 2 e^{\sigma} (\rho + p) \delta u_t + p \, \delta g_{tt}\;,
\end{align}
which completes the gravitational wave equation in Eq.\eqref{eq:gw02}. 

In this scenario, we model an oscillating star with a source \( \delta T_{tt} \). We assume a compact star at rest in the Weyl metric with radial oscillations, so \( \rho(t), p(t) \) oscillate around equilibrium. This means that the perturbation \( \delta T_{tt}(t) \) acts as a source of gravitational radiation in the perturbed Einstein equations, generating GWs, influenced by both the star's motion and the background geometry. In this sense, we have
\begin{align}\label{eq:tmunu01}
\delta T_{tt}(t, r) = \delta\rho(t, r) g_{tt}(r) + p(t, r) h_{tt}(t, r)\;,
\end{align}
and the ansatz
\begin{align}\label{eq:tmunu02}
\delta\rho(t, r) = \rho_0 \cdot \cos(\omega t) \cdot e^{-r^2/R^2}\;,
\end{align}
where \(\delta\rho(t, r)\) is a time-dependent perturbation and $\rho_0$s the amplitude of the density perturbation. Thus, the GW source is localized and periodic as the Weyl background modifies the propagation. Hence, we can simulate a star-induced waveform as
\begin{align}\label{eq:waveform}
h(t) = A_0 \cos(2\pi f_0 t + \phi)e^{\sigma(r_0, z_0) - \lambda(r_0, z_0)}\nonumber \\\hspace{1cm}\;\;\left(1 + \epsilon \cos(2\pi f_{\text{star}} t)\right) 
\end{align}
where \( f_0 \) is the base GW frequency, \( f_{\text{star}} \) is the frequency of star oscillation, and \( \epsilon \ll 1 \) is the modulation amplitude that keeps in the linear regime.

As discussed in earlier sections, the solutions of Eqs.\eqref{eq:sigmabessel} and \eqref{eq:lambdabessel} constitute the best case to examine. From the Weyl metric, we know that \(\sigma(r,z)\) dominates the gravitational redshift, time dilation, and the Newtonian-like potential (appears in \( g_{tt} \)), and it is the most relevant function for geodesics and test particle motion. On the other hand, \(\lambda(r,z)\) denotes a correction term that ensures that the field equations are satisfied. But not only this. Since the solutions \(\sigma(r,z)\) and \(\lambda(r,z)\) are found from non-linear systems of equations (Eqs.\eqref{eq:01}, \eqref{eq:02} and \eqref{eq:03}), \(\lambda(r,z)\) correlates with \(\sigma(r,z)\) and contributes to the curvature of spatial slices. Thus, it modulates the amplitude of spatial contributions (e.g., stress-energy conservation), but not the core of GW dynamics, that is, it does not critically affect the gravitational waveform dynamics in our model.

In this modeling, we assume a star-induced GW model where the source is localized, and GWs propagate through the exterior vacuum. To maintain causal structure, we consider \( \lambda(r, z) \) only for \( r > R_* \) that matches the domain of validity of the vacuum solution.

\subsection{Numerical analysis and implications}
Here, we summarize the main steps of our code implementation and physical implications. We start our numerical analysis by creating a \texttt{Python} code capable of simulating a scalar wave \( h_{tt}(r,z,t) \) propagating in a curved spacetime influenced by the two potentials \(\sigma(r,z)\) and \(\lambda(r,z)\). The \(\sigma(r,z)\) potential of Eq.\eqref{eq:sigmabessel} is a damping function with exponential decay along the \( z\) coordinate and presents a radial dependence with a Bessel function \( J_0 \). On the other hand,\(\lambda(r,z)\) is a potential active only outside a central star of radius \( R_* = 0.5 \), involving \( J_0 \) and \( J_1 \) Bessel functions. The evolution follows a modified wave equation with redshift effects 
of Eq.\eqref{eq:waveform}. We assume a Gaussian-shaped source oscillating at frequency \( \omega = 300 \).  

\begin{figure}[t!]
\centering
\mbox{    
\subfloat{\includegraphics[width=3.1in, height=2.8in]{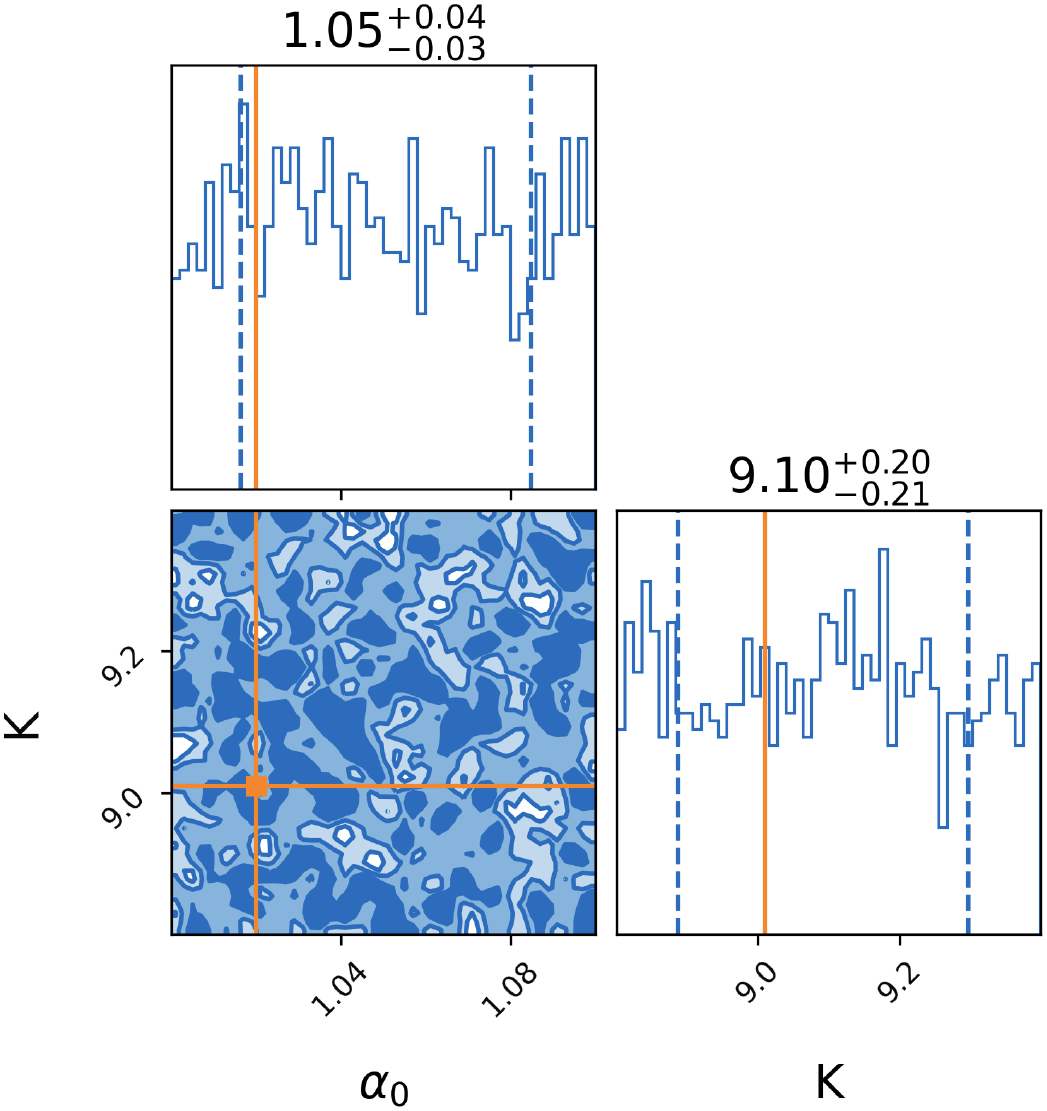}}
}
\caption{Corner plot of the parameters $(\alpha_0, K)$. Top panel shows the marginalized posterior for \( \alpha_0 \)
and the right panel shows the marginalized posterior for \( K \). Each of these histograms shows a solid orange vertical line (median values) and the two dashed blue vertical lines pinpoints the bounds of the $68\%$ of credible intervals~(CI).}
\label{fig:contours}
\end{figure}

\begin{figure*}[t!]
\centering
\mbox{    
\subfloat{\includegraphics[width=6.5in, height=2.2in]{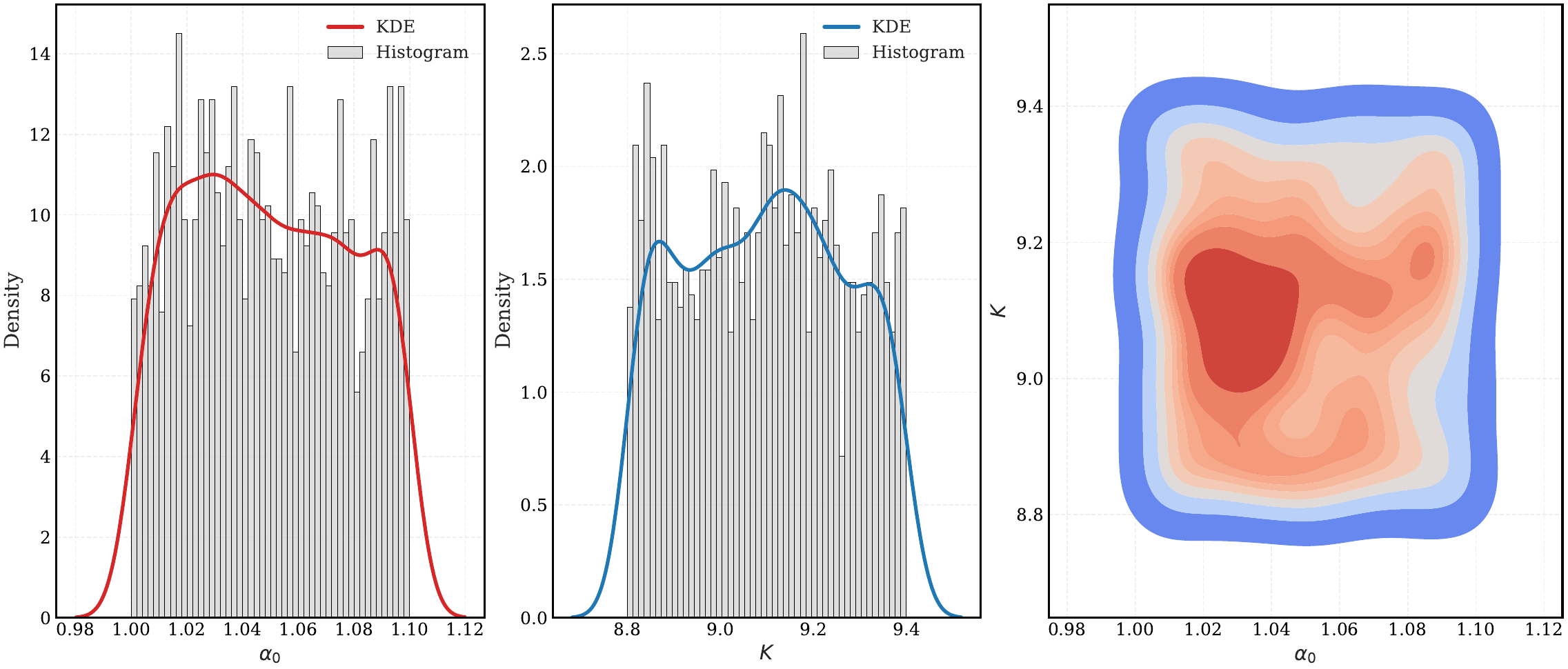}}
}
\caption{KDE smoothing of the posteriors of parameters. Left plot shows posteriors of $\alpha_0$, while the middle panel shows the posteriors of $K$. The right panel shows smoothed 2D contours for joint distribution in the plane \((K-\alpha_0)\).}
\label{fig:kde}
\end{figure*}

We impose the initial conditions with a Gaussian pulse centered at \( (r_0=0, z_0=1) \) with width \( \sigma_s = 0.2 \), modulated by \( \cos(\omega t) \).  The grid spans \( r \in [0, 2] \), \( z \in [0, 2] \), and \( t \in [0, 1.01] \) with steps \( dr, dz, dt \).  Using the Finite-Difference Scheme~(FDS), we have the time-stepping that uses a second-order ``leapfrog'' method as  
\begin{align}
h_{tt}^{n+1} = 2h_{tt}^n - h_{tt}^{n-1} + dt^2 \left[ e^{-2(\lambda - \sigma)} \nabla^2 h_{tt}^n + \text{source} \right],
\end{align}
where \textit{n} represents the current time step index in the discretized time evolution of the wave equation. The spatial derivatives are approximated via centered differences (e.g., \( \partial_r^2 h_{tt} \approx \frac{h_{tt}(r+dr) - 2h_{tt}(r) + h_{tt}(r-dr)}{dr^2} \)).  We point out that the code is efficient for simple explicit FDS, but relies on \texttt{NumPy} + minimal \texttt{SciPy}. The boundary conditions (Neumann conditions) enforce a zero gradient at all boundaries, e.g. \( h_{tt}(0,z) = h_{tt}(dr,z) \).  For numerical stability, we use the Courant–Friedrichs–Lewy (CFL) condition \( dt \lesssim \min(dr, dz)/c \), where \( c \) is the effective wave speed.   As a result, the output is a time series of \(h_{tt}\) that captures the wave evolution under the defined potentials. For all simulations, we fix the code with parameters $(\alpha_0, K)$ by means of the \texttt{Bilby} ~\cite{bilby} module. \texttt{Bilby} is a python package known for Bayesian parameter estimation, often used in gravitational-wave data analysis. Synthetic data is generated by \texttt{Bilby} for testing or injection into simulated noise (Gaussian or real detector noise). Then, it passes to \texttt{Dynesty}~\cite{Speagle_2020} sampler to explore the parameter space to compute Bayesian posteriors and evidence efficiently. We define the initial parameter space with priors uniformly distributed  priors $\{ \alpha_0:(0.9, 1.5), K:(8.0, 10.0)\}$.  Over time, \texttt{Dynesty} focuses on areas where the parameters best explain the data around the ``true'' parameter values. Outside this range, we verified that we only obtain noise signs or nonphysical patterns.  The results are shown in Fig.(\ref{fig:contours}), where the solid orange vertical line denotes the median value and the two dashed blue vertical lines denote the bounds of the $68\%$ of credible interval~(CI). The histograms show the marginalized posterior distribution for a single parameter. The peak indicates the most probable value, and the width reflects the uncertainty. In the two-dimensional contours, they show the joint posterior distributions between pairs of parameters, suggesting parameter degeneracies and correlations. 

\begin{figure*}[t!]
\centering
\mbox{
\subfloat[]{\includegraphics[width=3.3in, height=2.2in]{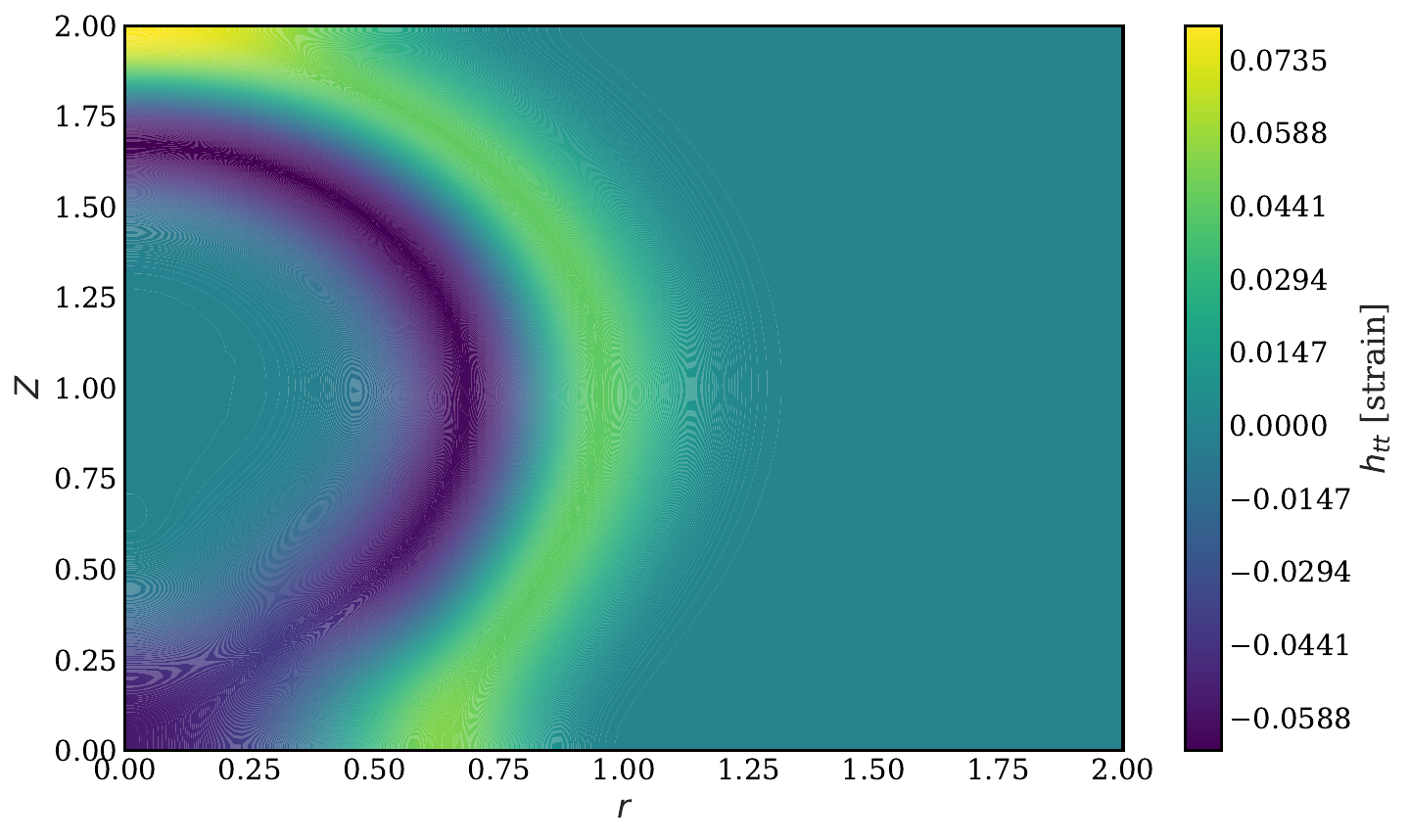}}\quad
\subfloat[]{\includegraphics[width=3.3in, height=2.2in]{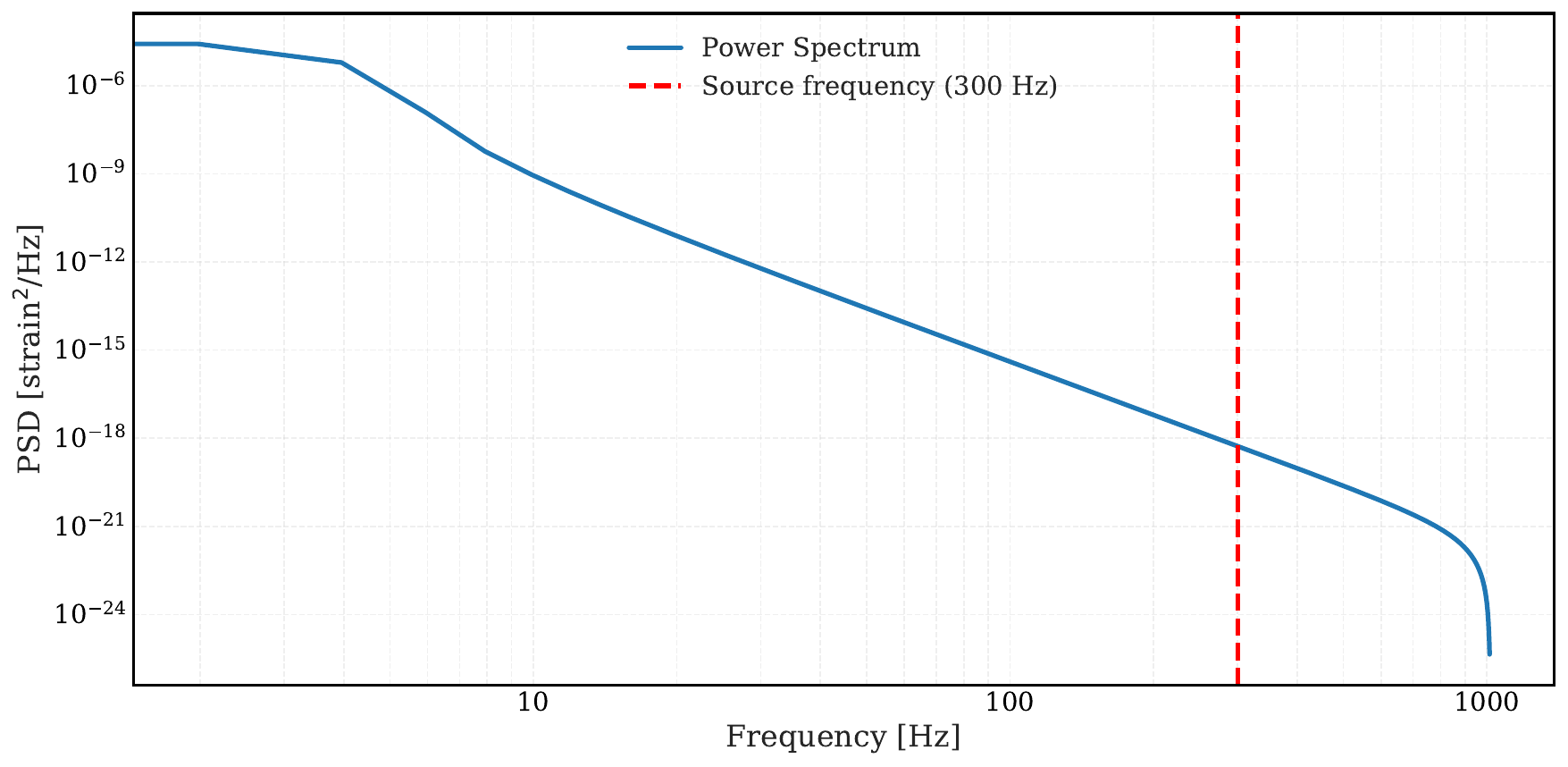}}
}
\mbox{
\subfloat[]{\includegraphics[width=3.5in, height=2.2in]{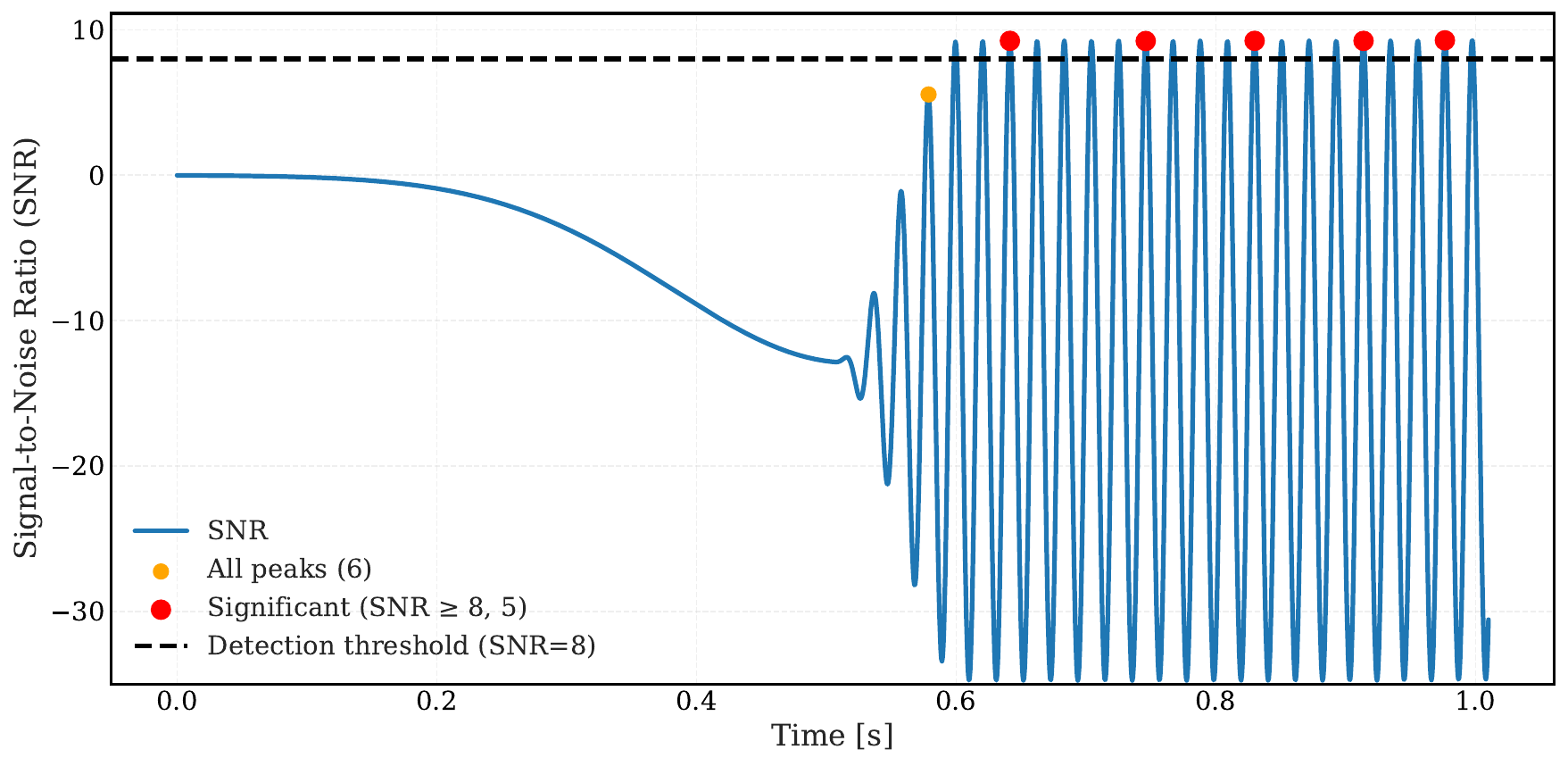}}\quad
}
\caption{In the upper left panel, it shows a spatial snapshot of a scalar or tensor field in plane $(r-z)$ of \( h_{tt} \) in a Weyl background. In the upper right panel, a frequency-domain plot of PSD from the waveform data. The bottom panel shows the Signal-to-Noise Ratio~(SNR) time series extracted from the simulated detector output. These peaks exhibit a regular temporal separation (\( \Delta t \sim 84 \, \text{ms} \)), suggesting a possible echo structure or quasi-periodic re-excitation of the source.} 
\label{fig:snr}
\end{figure*} 

In this case, the contours seem fairly circular, implying a weak correlation. This can be improved with post-processing KDE smoothing. In Fig.(\ref{fig:kde}), we present the KDE-smoothed posterior distributions for \( \alpha_0 \) and \( K \). The gray histograms show the raw sample distribution. The red and blue curves in the left and middle panels show the smoothed KDE fits for \( \alpha_0 \) and \( K \) parameters, respectively. Both parameters have well-peaked distributions, indicating strong constraints. The KDEs are smooth and unimodal, confirming no pathological sampling behavior and  \( K \) presents a slight asymmetry, which is expected for a non-Gaussian posterior. The right panel presents a 2D contour plot of \( \alpha_0 \) and \( K \) as a joint posterior distribution of those two parameters. Each contour line defines a region of parameter space where the joint probability density is constant. The filled contours correspond to different confidence levels: innermost contour indicates $68\%$ credible region, while outer contours point higher cumulative probability regions $(95\%, 99\%)$. The darker regions pinpoint where samples are dense, i.e., indicating higher probability. The contours now look more evident fairly circular, suggesting weak correlation between the parameters, implying that the model distinguishes the effects of \( \alpha_0 \) and \( K \) cleanly.

Fig.(\ref{fig:snr}) shows plots to analyze the wave propagation and detection. The waveform is generated from a physical model with parameters \( \alpha_0 = 1.05 \), \( K = 9.1 \), and \( R_{star}=0.5\). We adopt a Gaussian pulse with frequency \( \omega = 300 \) Hz that evolves in a 2D grid (\( r, z \)) over time. The left upper panel (a) shows a snapshot of the field \( \bar{h}_{tt}(r, z, t) \) at a fixed time slice \( t = 0.8 \)s, extracted halfway through the simulation. The spatial profile reveals an approximately axisymmetric structure, with oscillatory features centered along the axis origin \( r = 0 \), consistent with a compact source located at \( z_0 \). The field exhibits strong localization and decay at large radii, consistent with energy confinement and finite support of the source term \( \delta T_{tt} \). The interference patterns and lobes suggest the emergence of multipolar structure, hinting at higher harmonic contributions.

The color gradient represents the amplitude of \( \bar{h}_{tt} \), with warm colors (greens/yellows) indicating positive field values, and cool colors (blues/purples) indicating negative values. The sidebar provides a quantitative measure of the field that varies smoothly between \( \bar{h}_{tt} \approx -0.02 \) (dark blue) and \( \bar{h}_{tt} \approx +0.02 \) (bright yellow). The sharpness of these color transitions corresponds to the local gradient of the field, indicating regions of higher oscillatory intensity. The bright central band stretching horizontally near \( z = z_0 \) marks the primary location of the compact source. This corresponds to the localized energy density driving the oscillation. The field is most intense here, forming a central ``core'' with near-symmetric lobes above and below, consistent with an approximate quadrupolar mode. The relative weakness of the field away from the core, the rapid fall-off of color intensity—implies that the compact object is radiating inefficiently, or the curvature of the Weyl background may be absorbing, redshifting, or delaying part of the outgoing signal. On the other hand, the field is highly localized in \( r\), consistent with a compact stellar source that does not excite wide-angle radiation.

The right upper panel (b) shows the Power Spectral Density (PSD) that confirms the source frequency at 300 Hz, marked by a red dashed vertical line, falls in LIGO's sensitivity band, in the range $20 \ \text{Hz} \lesssim f \lesssim 1000 \ \text{Hz}$. The power spectrum displays a dominant peak around \( f \sim f_0 \), corresponding to the fundamental oscillation mode of the compact source. Sub-dominant peaks at higher harmonics reflect nonlinear mode coupling and echo-like features. The spectral shape is influenced by both the source structure and the Weyl curvature corrections to the wave propagation, with potential redshifting effects visible as asymmetries in the tail.
\begin{figure*}[t!]
\centering
\mbox{    
\subfloat{\includegraphics[width=0.9\linewidth]{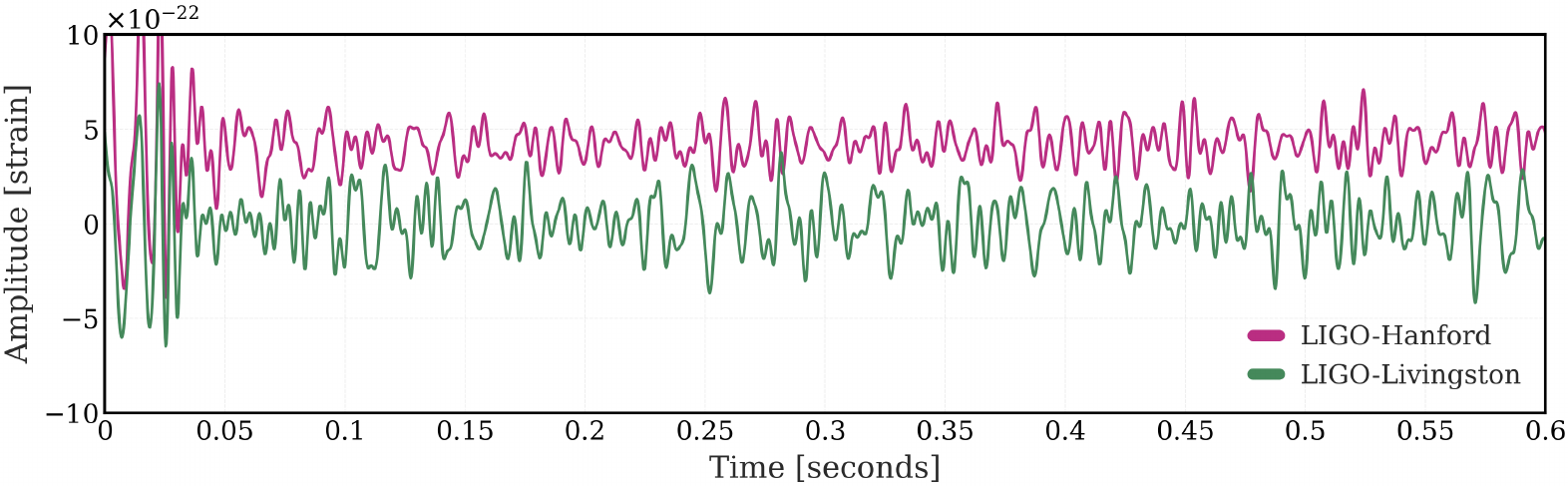}}
}
\caption{Gravitational wave strain data from the LIGO-Hanford~(H1) and LIGO-Livingston~(L1) detectors. It presents a longer sign duration larger than a typical BBH merger but consistent with an exotic compact object (ECO).}
\label{fig:ligo}   
\end{figure*}

This panel presents the frequency-domain representation of the gravitational signal \( \bar{h}_{tt}(t) \), as extracted from a distant point (typically far along the symmetry axis or in the wave zone). The vertical axis gives PSD, proportional to the square of the Fourier amplitude, while the horizontal axis shows the corresponding frequency (in units consistent with the simulation's normalization, e.g., \( M^{-1} \)). The power spectrum is sharply peaked around a single frequency \( f_0 \), with negligible sidebands or harmonics. This narrowband structure confirms that the signal is essentially monochromatic corresponding to a single, stable oscillation mode of the source. We did not identify any apparent nonlinear coupling, overtone excitation, or multi-mode interference in the wave zone. The clear single peak suggests that the simulation resolved the dominant oscillation mode well, and the extraction method is clean. This type of signal is ideal for gravitational wave template matching, allowing precise frequency-based inference. If this were a real detection, the peak frequency could be used to constrain the compact object's equation of state or background curvature effects due to the Weyl background. The lack of broadband features means that it is likely a quasi-stationary phase of the waveform, not the initial burst or final decay.

The third panel is show in the bottom figure (c) and exhibits the time-domain Signal-to-Noise ratio (SNR) recovered from matched-filtering with the custom Weyl-modulated waveform. The SNR sharply rises during the interval \( t \in [t_1, t_2] \), marking the arrival and peak amplitude of the GW signal. The subsequent decay is consistent with ringdown and tail contributions. It shows a five out of six peaks that exceed the detection threshold $SNR\geq 8$ analyzed for false alarm rates (FAR). The main SNR curve (blue) shows fluctuations, with detected peaks marked as orange dots.  Peaks exceeding SNR=8 (red dots) are flagged as significant detections. A dashed black line indicates the detection threshold (SNR=8). We obtain a SNR around (9.2–9.3) that is typically considered marginal but still detectable, and 
FAR effectively zero, which means that none of the peaks can be explained by Gaussian noise under the detection model. This SNR structure is not typical of binary black hole mergers or standard neutron star inspirals, suggesting an exotic compact object, like a boson star or quark star with internal oscillation burts.

In Fig.(\ref{fig:ligo}), it shows a comparison with the time-domain gravitational wave signals as recorded by the LIGO Hanford (H1) and Livingston (L1) detectors. We use \texttt{Python} packages \texttt{Gwosc}~\cite{Abbott:2021:LIGO12,Abbott:2023:LIGO3} for database and \texttt{GWpy}~\cite{gwpy} for simulating time series data from H1 and L1 detectors. Then, we can inject our Weyl waveform into LIGO data around the time of GW150914. The panel presents the normalized signal amplitude versus time in seconds. In H1~(cyan line) and L1~(green line) detectors, the signals display a similar waveform indicating the detection of the same astrophysical event as well a slight time delay. The characteristic chirp is also visible, which it strongly suggests to a real gravitational wave event, not noise. For an astrophysical fingerprint the middle part of the waveform shows better indications of physical viability of the sign, since it has a gradual increase in frequency (oscillations get closer together), and a rise in amplitude (waveform grows taller), suggesting an inspiraling binary system in the final seconds before merger or a black-hole-like. The early parts of the sign is weaker, possibly in the noise floor and late parts (after the peaks) is often ringdown or noise-dominated as well. This is longer than a typical BBH merger $\sim (0.2 s)$ but again consistent with an ECO like a boson star or quark star, especially one undergoing post-merger oscillations.

\begin{table}[t!]
\centering
\caption{Bayesian evidence and model fitting results}
\begin{tabular}{lcc}
\hline
\textbf{Metric} & \textbf{Value} \\
\hline
Number of samples & 1.168 \\
Log noise evidence ($\ln \mathcal{Z}_{\text{noise}}$) & -3.129 \\
Log evidence ($\ln \mathcal{Z}$) & -3.188 $\pm$ 0.008 \\
Log Bayes factor ($\ln \mathcal{B}$) & 0.060 $\pm$ 0.008 \\
Best Quark model amplitude & 0.113 \\
Best Quark model shift & 0.100 \\
Quark model score & 0.048 \\
Weyl vs Boson score & 0.103 \\
Weyl vs Quark score & 0.048 \\
\hline
\end{tabular}
\label{tab:evidence}
\end{table}

\begin{table}[t!]
\centering
\caption{Cross-validation model scores}
\begin{tabular}{ccc}
\hline
\textbf{Fold} & \textbf{Weyl vs Boson} & \textbf{Weyl vs Quark} \\
\hline
1 &  0.210 & 0.135 \\
2 &  0.141 & 0.071 \\
3 &  0.103 & 0.048 \\
\hline
\textbf{Average} &  0.151 $\pm$ 0.044 & 0.085 $\pm$ 0.037 \\
\hline
\end{tabular}
\label{tab:cv_scores}
\end{table}

Next, we run a Bayesian model selection code between Boson star and Quark star with \texttt{Bilby}. We obtain the results shown in Tables (\ref{tab:evidence}) and (\ref{tab:cv_scores}). The values presented in Table (\ref{tab:evidence}) show the overall evidence for the model against the noise and other models. The log-Bayes factor is close to zero, suggesting that there is not a strong preference for one model over another in terms of evidence. However, the relatively small error in the log-evidence indicates that the model has a stable likelihood estimation.

In Table (\ref{tab:cv_scores}), we apply the cross-validation method for evaluating the performance of the model by partitioning the data into multiple subsets (folds). In our case, we study the accuracy and robustness of our models (Weyl, Boson, Quark) by analyzing how well they generalize across different data splits. The Boson vs Weyl scores across the three folds (0.210, 0.141, 0.103) show a consistent trend, but they also indicate some variability. Fold 1 gives the highest score, suggesting that in some subsets of the data, the Weyl model fits the Boson waveform better. In fold 2 and fold 3, the score drops, suggesting a less optimal match. Similarly, the Quark vs Weyl scores (0.135, 0.071, 0.048) show a decline across folds, with Fold 3 giving the lowest score, indicating that the Weyl model consistently performs worse in matching the Quark waveform. This suggests that, on average, the Weyl model is better matched with the Boson waveform than the Quark waveform, as expected based on our earlier analysis.

\section{Final remarks}
In this work, we have studied vacuum solutions of the Weyl metric. Besides of being a very known metric in literature, we have presented new radial, oscillatory, logarithmic and exponential solutions. By means of the Hamilton-Jacobi approach, we obtain the general integrals of motion and the effective potential, accordingly. Investigating the radial solution, we have found that it generalizes the Curzon-Chazy solution with the appearance of additional integration constants. Analyzing its causal structure, it has a singularity at the origin \(r=z=0\), and for $k2 < 0, k3 < 0$ with $k_2\neq k_3$, it may support realistic disk formation and shadow structure, since its ISCO is reasonable for efficient accretion. 
The second part of solutions give three cases depending on the type of sign of the integration constant $K$. For $K=0$, we have the logarithmic solution well-explored in literature. The novelty lies in the cases $K>0$ and $K<0$. We have chosen the exponential case with $K>0$ that presents a more stable causal structure and regularity at origin and asymptotic domain. The oscillatory solution $K<0$ has revealed to be nonphysical.

Taking the exponential solution, we have modeled GW emission from a compact star undergoing oscillations in a static, axisymmetric Weyl background. The waveform has been constructed as a modulated sinusoid where \( f_0 \) corresponds to the dominant GW emission frequency typically associated with the star's quadrupolar f-mode~\cite{Kokkotas1999} and \( f_{\text{star}} \) represents a lower-frequency stellar oscillation (e.g., a radial or g-mode). The modulation amplitude \( \epsilon \ll 1 \) ensures a linear regime, allowing interpretation as amplitude modulation of the primary GW signal due to internal fluid dynamics \cite{Ferrari2003}.

In the modeling of GW wave emission from a compact star in a static, axially symmetric Weyl background, the gravitational potential \( \sigma(r, z) \) governs the dominant physical effects such as redshift and time dilation. Since the metric function \( g_{tt} = -e^{2\sigma} \) determines the causal structure and directly influences wave propagation, \( \sigma \) is primarily responsible for shaping the dynamics of the emitted signal. The companion potential \( \lambda(r, z) \), while required for the full metric to satisfy Einstein’s vacuum equations, enters only through spatial curvature corrections and volume elements. In this context, its influence is subleading. We define \( \lambda(r, z) \) only for \( r > R_* \), consistent with the exterior region where the Weyl metric is valid. This restriction reflects the physical assumption that the interior of the star is governed by a different (non-vacuum) solution, and it allows us to cleanly separate the source region from the propagation zone. Thus, including \( \lambda \) for \( r > R_* \) serves to maintain mathematical completeness without significantly affecting the gravitational waveform's leading-order features.

The exponential factor \( e^{\sigma - \lambda}\) encodes the effect of the Weyl background geometry, evaluated at the source location \( (r_0, z_0) \), and modifies the observed amplitude as a curvature-dependent scaling factor \cite{Stephani2003}. This captures the impact of the surrounding spacetime on the star’s emission profile without altering the core dynamics of the fluid source. While a spherically symmetric perfect fluid does not radiate in general relativity \cite{Maggiore2008}, the presence of a time-varying quadrupole — as implied by \( \cos(2\pi f_0 t) \) — allows for physical GW generation. The modulation term serves as a phenomenological proxy for coupling between internal fluid modes and spacetime curvature. As a result, our finding suggest an astrophysical event consistent with weak, long-duration, or narrowband signals (SNR 9.2-9.3) close to the threshold of detectability indicating an exotic Boson star bursts. The Weyl metric indicates a flexible framework to describe a range of compact objects, including boson stars. The model may generalized by incorporating additional terms for higher-order curvature effects or more complex forms of scalar field dynamics, potentially using a hybrid approach combining Weyl geometry with scalar field dynamics.

\begin{acknowledgements}
AJSC acknowledges Conselho Nacional de Desenvolvimento Cient\'{i}fico e Tecnologico (CNPq) for the partial financial support for this work (Grant No. 305881/2022-1) and Fundação da Universidade Federal do Paraná (FUNPAR, Paraná Federal University Foundation) through public notice 04/2023-Pesquisa/PRPPG/UFPR for the partial financial support (Process No. 23075.019406/2023-92), and the financial support of the NAPI ``Fenômenos Extremos do Universo" of Fundação de Apoio à Ciência, Tecnologia e Inovação do Paraná.

This research has made use of data or software obtained from the Gravitational Wave Open Science Center (gwosc.org), a service of the LIGO Scientific Collaboration, the Virgo Collaboration, and KAGRA. This material is based upon work supported by NSF's LIGO Laboratory which is a major facility fully funded by the National Science Foundation, as well as the Science and Technology Facilities Council (STFC) of the United Kingdom, the Max-Planck-Society (MPS), and the State of Niedersachsen/Germany for support of the construction of Advanced LIGO and construction and operation of the GEO600 detector. Additional support for Advanced LIGO was provided by the Australian Research Council. Virgo is funded, through the European Gravitational Observatory (EGO), by the French Centre National de Recherche Scientifique (CNRS), the Italian Istituto Nazionale di Fisica Nucleare (INFN) and the Dutch Nikhef, with contributions by institutions from Belgium, Germany, Greece, Hungary, Ireland, Japan, Monaco, Poland, Portugal, Spain. KAGRA is supported by Ministry of Education, Culture, Sports, Science and Technology (MEXT), Japan Society for the Promotion of Science (JSPS) in Japan; National Research Foundation (NRF) and Ministry of Science and ICT (MSIT) in Korea; Academia Sinica (AS) and National Science and Technology Council (NSTC) in Taiwan.

\end{acknowledgements}

\bibliographystyle{spphys} 
\bibliography{static_BHweyl}

\end{document}